\documentclass[%
aip,
cha,
amsmath,amssymb,
reprint,
numerical,%
]{revtex4-2}

\usepackage{graphicx}
\usepackage{dcolumn}
\usepackage{bm}

\usepackage[utf8]{inputenc}
\usepackage[T1]{fontenc}
\usepackage{mathptmx}
\usepackage{etoolbox}
\usepackage[dvipsnames]{xcolor}
\usepackage{framed}

\usepackage{ulem}

\makeatletter
\def\@email#1#2{%
 \endgroup
 \patchcmd{\titleblock@produce}
  {\frontmatter@RRAPformat}
  {\frontmatter@RRAPformat{\produce@RRAP{*#1\href{mailto:#2}{#2}}}\frontmatter@RRAPformat}
  {}{}
}%
\makeatother
\begin{document}

\title{
Basin of attraction organization in infinite-dimensional delayed systems: a stochastic basin entropy approach}
\author{Juan Pedro Tarigo}
\affiliation{ 
Instituto de Física, Facultad de Ciencias, Universidad de la República, Igua 4225, 11400 Montevideo, Uruguay
}%
\author{Cecilia Stari}%
\affiliation{ 
Instituto de Física, Facultad de Ingeniería, Universidad de la República, Julio Herrera y Reissig 565, 11300 Montevideo, Uruguay
}%
\author{Arturo C. Mart\'i}
\email{marti@fisica.edu.uy}

 \homepage{http://www.fisica.edu.uy/~marti}
\affiliation{ 
Instituto de Física, Facultad de Ciencias, Universidad de la República, Igua 4225, 11400 Montevideo, Uruguay
}%
\date{\today}

\begin{abstract}
  The Mackey-Glass system is a paradigmatic example of a delayed model
  whose dynamics is particularly complex due to, among other factors,
  its multistability involving the coexistence of many periodic and
  chaotic attractors. The prediction of the long-term dynamics is
  especially challenging in these systems, where the dimensionality is
  infinite and initial conditions must be specified as a function in a
  finite time interval. In this paper we extend the recently proposed
  basin entropy to randomly sample arbitrarily high-dimensional
  spaces. By complementing this stochastic approach with the basin
  fraction of the attractors in the initial conditions space we can
  understand the structure of the basins of attraction and how they
  are intermixed.  The results reported here allow us to quantify the
  predictability giving us an idea about the long-term evolution of
  trajectories as a function of the initial conditions.  The tools
  employed can result very useful in the study of complex systems of
  infinite dimension.
\end{abstract}

\maketitle

\textbf{In many real systems, ranging from climate to brain dynamics,
  long-term predictability is challenging due to chaotic solutions and
  high sensitivity to initial conditions. Multistable systems, in
  particular, exhibit unpredictable behavior that varies with control
  parameters.  In these cases it is more a question of which attractor
  the system tends towards and not so much the value of the (local)
  Lyapunov exponents.  Basin entropy is a new tool designed to measure
  predictability by partitioning initial condition spaces and
  analyzing their long-term dynamics. In time-delayed systems, and in
  particular in the Mackey-Glass system, the complexity increases as
  the system's dimensionality becomes arbitrarily large, complicating
  the exploration of initial conditions. In earlier work, we
  demonstrated that basin entropy is effective even in simple
  time-delayed bistable systems for studying multistability and
  predicting bifurcations. We extend this to the Mackey-Glass model,
  known for its rich dynamics and multiple attractors, showing that a
  modified, stochastic version of basin entropy in conjunction with
  the number of attractors and the relative volumen they occupy
  effectively characterizes the system's behavior.  }

\section{Introduction}
In nonlinear systems the predictability of the long-term dynamics is
strongly limited by the presence of chaotic solutions and the high
sensitivity to initial conditions.  Particularly in multistable
systems the predictability is not at all uniform and strongly depends
on the control parameters.  In this situation, the analysis of basins
of attraction plays a crucial role \cite{menck2013basin,leng2016basin,
  schultz2017potentials,rakshit2017basin,Gelbrecht_2021,datseris2023framework}.
Basin entropy is a recently proposed tool to quantify the
predictability of long-term dynamics in these systems
\cite{daza2016basin,daza2017chaotic}. Its main idea consists in
dividing the space of initial conditions into regular partitions and
studying within each one the long time evolution of the dynamics. This
tool was successfully applied in several systems such as the analysis
of fractal boundaries \cite{gusso2021fractal}, to classify different
kinds of basins as Wada, riddled or intermingled
\cite{daza2022classifying}, or to explore bifurcations
\cite{daza2023unpredictability,wagemakers2023using}.  Recently, an
innovative strategy for characterizing the complexity of basins in
dynamical systems using convolutional neural networks, demonstrating
their superior efficiency over traditional methods in analyzing
multiple basins of attraction \cite{valle2024deep}.  In case of
complex networks naturally composed of many nodes the dynamic
stability of the basins of attraction also plays an important role
\cite{halekotte2020minimal, halekotte2021transient}.

In systems with time delay the additional complication arises that the
dimension of the system is arbitrarily large and the space of initial
conditions cannot be exhaustively explored
\cite{gros_2019,radons_2019}. In a previous work
\cite{tarigo2024basin} we considered one of the simplest time-delayed
systems, a bistable system with a delay term, which exhibits complex
dynamics \cite{erneux2009applied}. We showed that even in this system
the basin entropy is a valid tool to study multistability and can be
an indicator of the occurrence of bifurcations. Here, we consider the
Mackey-Glass (MG) model, a time-delayed system that exhibits
particularly rich dynamics with regions impacted by the existence of
many attractors and we show that the basin entropy modified to
stochastically partition the initial condition space it is useful to
characterize the dynamics of the system.  To complement the results
obtained with this tool we also consider the basin fraction occupied
by the attractors \cite{menck2013basin,leng2016basin} in the space of
initial configurations and the analysis of cross-sections obtained for
particular families of initial condition functions. Our approach is
based on the joint use of these three tools.

The Mackey-Glass model initially proposed for the study of
physiological systems is a paradigmatic example of a time-delayed
system that exhibits very complex dynamics
\cite{mackey1977oscillation}.  Its dynamics is characterized by
multistability that includes the coexistence of different attractors,
fixed points, limit cycles or chaotic attractors
\cite{junges2012intricate,amil2015exact,amil2016electronically}.  In
Ref. \onlinecite{junges2012intricate}, J. A. C. Gallas \textit{et al}
showed that the dynamics exhibit periodic or aperiodic oscillations
where peaks appear and disappear due to continuous deformations as the
control parameters vary. Bifurcation diagrams revealed the existence
of a complex structure of period-doubling or peak-adding bifurcations
leading to chaotic windows. They also obtained stability diagrams in
terms of control parameters and effective delayed showing complex
mosaics of periodic and chaotic regions.  The variation of the delay
time allowed them to observe some kind of \textit{lethargy} with which
the system responds to the change from one to many degrees of freedom,
demonstrating that the emergence of high dimensionality does not
instantaneously alter the dynamics of the system.  This observation
challenged the general validity of some analytical methods previously
considered in the literature.

In general, as a consequence of time delay, it is required to specify
a continuous function of initial conditions over a finite interval to
determine its evolution and dimensionality results to be arbitrarily
large. In a previous work we designed an electronic circuit with
feedback that simulates the MG system and showed that the experimental
results obtained with the circuit reproduce the behavior of the model
over a wide range of parameters \cite{amil2015exact}. This
implementation served to show that, taking arbitrary families of
initial conditions, in general different attractors coexist for the
same value of the control parameters. These attractors can be either
fixed points, or periodic oscillations with equal amplitude maxima but
different orderings, or chaotic attractors
\cite{amil2015organization}. Additionally, hysteresis loops were
identified by varying the delay parameter in bifurcation diagrams.
Subsequently, we focused on precisely quantify the multistability in
this system \cite{tarigo2022characterizing}. With this objective, we
selected representative initial condition functions and systematically
explored the parameter space counting the number of different
attractors. In this way, we identified the regions in which more
attractors coexist and assessed the impact of multistability.  In the
present work we go deeper into this subject and use the basin entropy,
basin fraction, number of attractors and the way in which they are
intermingled to analyze the dynamics and predictability of the system.

\section{The Mackey-Glass delayed model} \label{sec:model}

The Mackey-Glass model is a time-delayed differential equation widely
used to study complex dynamic behaviors in systems exhibiting chaos
and multistability \cite{mackey1977oscillation}. Originally proposed
by Michael Mackey and Leon Glass in 1977 to describe physiological
control systems, such as blood cell regulation, the model has since
been applied in various fields including biology, physics, and
engineering. The equation characterizes how the rate of change of a
variable, such as the concentration of a substance, depends not only
on its current state but also on its state at some previous time. This
time delay introduces infinite-dimensional phase space, making the
Mackey-Glass model a prototypical example for investigating
time-delayed systems. Its ability to generate rich dynamics, including
periodic, quasi-periodic, and chaotic solutions, makes it a valuable
tool for exploring fundamental principles of nonlinear dynamics and
chaos theory.

Let us denote $x(t)$ the concentration of a certain blood component at
time $t$.  The dynamics of the system can be described by means of the
Mackey-Glass model in terms of the balance between a nonlinear
production term and a decay term. Using dimensionless variables the
model equation can be written as
\begin{equation}
\frac{dx}{dt}=\alpha\frac{x_{\Gamma}}{1+x_{\Gamma}^{n}}-x
\label{eq:MGlinda}
\end{equation}
where $x_{\Gamma}= x(t - \Gamma)$ is the delayed state variable,
$\Gamma$ plays the role of an effective delay.  In addition, $\alpha$
determines the scale of the production term while the exponent $n$ is
related to the width of the function. Equation \ref{eq:MGlinda} is a
delayed differential equation its dimensionality is infinite and to
determine the evolution it is necessary to know a initial condition
function over a finite interval $(-\tau,0]$.  The dynamical properties
  of time-delayed models differ from those of ordinary models, and
  their study presents important challenges.

In a previous work, we designed an electronic circuit that reproduces
the dynamics of the Mackey-Glass model \cite{amil2016electronically}.
In general terms, this circuit consists of a block that models the
nolinear production function and another, known as the bucket brigade
device, that stores and updates the values of the variables to
simulate the delay term. This circuit can be modelled by an explicit
discrete-time equation, which in the continuous limit coincides with
the MG model.  The numerical results obtained with both the discrete
map of dimension $N$ and integration methods demonstrate significant
agreement with those obtained experimentally.

To obtain the discrete-time evolution equation, we consider a small
time interval $\Delta t$ and that the delay verifies $\Gamma= N \Delta
t$, and we integrate Eq.~\ref{eq:MGlinda} between $t$ and $t + \Delta
t$, assuming that the \textit{delayed} variable takes a constant
value, we can obtain an equation for the time evolution of the system
during a small time interval.\begin{equation} x(t +\Delta t)= \left(
  x(t) - \frac{\alpha x_{\Gamma}} {1+x_{\Gamma}^n} \right) e^{- \Delta
    t} + \frac{\alpha x_{\Gamma}} {1+x_{\Gamma}^n }
 \label{eq:integrada}
\end{equation}
This equation serves as a basis for obtaining a map describing the
time evolution of the system by means of a continuous variable taking
discrete time values.  Defining $x_i = x( i \Delta t)$, we can write a
delayed one-dimensional map. In such a case to obtain $x_{i+1}$, it is
necessary to know the value of the variable in the previous instants
$x_i, x_{i-1} , ... x_{i-N}$.  A one-dimensional map where the current
variable depends on $N$ previous times is equivalent to an
instantaneous map of dimension $N$
\cite{masoller2003synchronization}. For simplicity we keep the same
notation for the variable $x$ in both cases resulting
\begin{align}
    \begin{split}
        x_1(t + 1) &= e^{-\frac{\Gamma}{N}}x_1(t) + \Big(1 - e^{-\frac{\Gamma}{N}}\Big) \alpha \frac{x_N(t)}{1 + x^n_N(t)} \\
        x_i(t + 1) &= x_{i - 1}(t) \quad \forall i = 2,...,N.
        \label{eq::map}
    \end{split}
\end{align}
To sum up, in this approximation the evolution of  the continuous-time variable $x(t)$, Eq.~\ref{eq:MGlinda}, is approximated by $N$  continuous variables at discrete times, $x_i(t)$, $\forall i=1,...N$ whose dynamics is given by Eqs.~\ref{eq::map}.

The dynamics of $x_1(t)$ given by Eq.~\ref{eq::map} reproduces the
time evolution of the continuous variable by taking the value at the
delayed time from the $x_N(t)$. The remaining units $x_i(t)$ for $i>1$
simply transfer their value to the adjacent variable at each time
step.  It can be seen that \eqref{eq::map} converges to
\eqref{eq:MGlinda} in the continuous limit $\Delta t = \Gamma/{N}
\rightarrow 0$ and that $N$ turns out to be the number of time steps
for one delay unit $\Gamma$. The state space of \eqref{eq::map} is
$N$-dimensional. This discretization of \eqref{eq:MGlinda} presents
the advantage of being exact in time \cite{amil2015exact} and faster
to compute than other conventional methods for numerically solving
DDEs. There are four control parameters in the discrete map, $n$ and
$\alpha$ determine the shape of the production term, while $N$ and
$\Gamma$ define the time delay.  In the following sections we are
going to use $n = 4$ and $N = 396$ unless otherwise is stated.  These
generic values coincide with some of those used in the experimental
and numerical results reported in the
Refs.~\cite{amil2016electronically,amil2015organization} and were
chosen to facilitate comparison.

\section{Stochastic basin entropy in high dimensional systems} \label{sec:be}
We begin by reviewing the definition of basin entropy for a
non-delayed dynamical system with $N_A$ coexisting attractors
\cite{daza2016basin,daza2017chaotic,daza2023unpredictability}. To
compute the basin entropy, we consider a partition of the phase space
into $N_b$ boxes of linear size $\varepsilon$ and examine which
attractor each point in each box evolves towards. We sample a large
number, $L$, of trajectories much greater than the number of possible
attractors inside box $i$ ($L \gg m_i$). This allows us to estimate
the probability, $p_{ij}$, of reaching the attractor $j$ starting from
an initial condition in box $i$, with the probabilities normalized so
that $\sum _j^{m_i} p_{ij}=1 $. The basin entropy is then defined as
\begin{equation}
S_b = - \frac{1}{N_b} \sum_{i=1}^{N_b} \sum_{j=1}^{m_i} p_{ij} \log p_{ij}
\label{eq::Sb}
\end{equation}
which can be identified as the Shannon entropy.  Given this definition
of the basin entropy and assuming all attractors inside box $i$ are
equiprobable ($p_{ij} = 1/m_i\, \forall j$) the uncertainty exponent
can be extracted of studying the relation betwen $S_b$ and the size of
the box $\varepsilon$ according to the method given in
Ref.~\onlinecite{daza2016basin}.

As the dimension of the phase space grows the algorithm in
\eqref{eq::Sb} becomes very computationally expensive and is not worth
using it for high dimensional systems. This problem arises due to the
fact that all the phase space must be explored in order to calculate
the basin entropy.  Because basin entropy is an average of the
quantity $p_{ij} \log p_{ij}$ over the entire phase space, it is
reasonable to postulate, following an approach somewhat similar to the
Monte Carlo method, that it will converge to the average value by
randomly sampling an adequate number of boxes. This method allows for
a much less computational effort provided that the basin entropy
converges rapidly to the average value.

To compute the basin entropy, rather than covering the entire space of
initial conditions with regular grids, we randomly select both the
center of each box and the initial conditions within it. For the
chosen parameter values, the relevant range for the initial conditions
lies between $0 + \varepsilon/2$ and $2 - \varepsilon/2$, since
negative values of $x$ have no physical significance, and larger
values will eventually relax to smaller ones. Therefore, the centers
of the boxes are randomly selected using a uniform distribution over
this interval, ensuring that all initial conditions fall within the
region of interest. In this work we use $\varepsilon = 0.3$ for all
calculations of the basin entropy and found similar results for other
values of $\varepsilon$. Next, we sample initial conditions within
each box, also following a uniform distribution, compute their
trajectories, and determine which attractor they evolve towards.

\begin{figure}[tb]
\centerline{\includegraphics[width=\columnwidth]{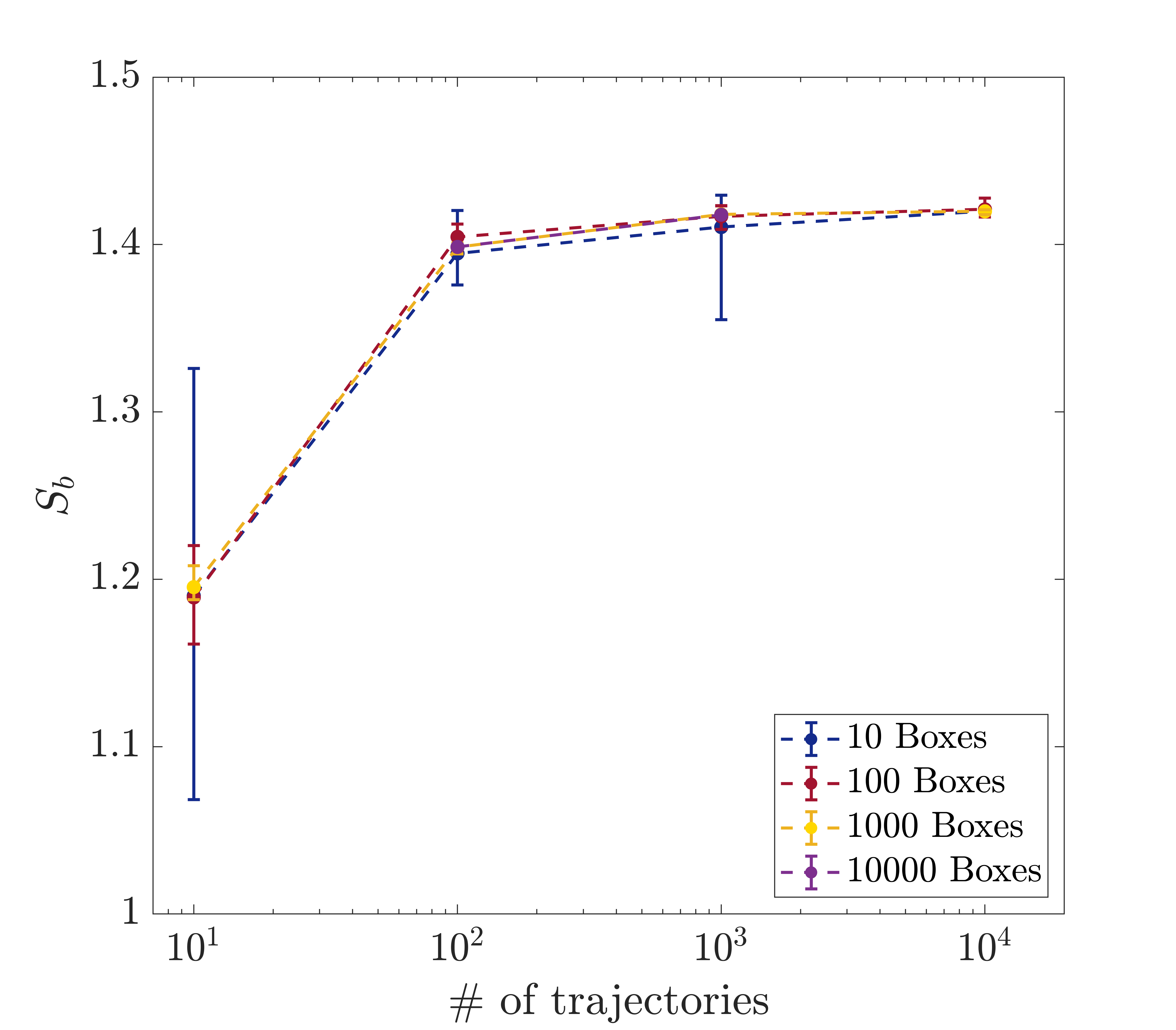}}
\caption{Convergence of basin entropy as a function of the number of
  initial conditions sampled per box for different number of boxes
  (indicated with colors in the legend box). For each point 10
  iterations of the basin entropy were averaged. The symbols indicate
  the average value while the error bars denote the minimum and
  maximum values found. All points correspond to the same parameter
  values of the model, $\alpha = 5$, $\Gamma = 18$ and $n = 8$.}
\label{fig_convergence}
\end{figure}

To determine the optimal number of boxes and trajectories, in
Fig.~\ref{fig_convergence} we calculated the basin entropy of the
system for different number of boxes and different number of
trajectories per box. We can see that as the number of trajectories
per box increases, the basin entropy starts to converge to a precise
value. Also increasing the number of boxes taken, decreases the
dispersion of the basin entropy around the mean value. We followed
this procedure for other values of the parameters of the system,
finding that the basin entropy converges before $100$ boxes and $1000$
trajectories per box in every case. In sections \ref{sec:results:qm}
and \ref{sec:results:bf} we use $100$ boxes and $1000$ trajectories
per box to calculate basin entropy.

\section{Visualizing high-dimensional sate spaces} \label{sec:resutls:ps}

\begin{figure}[tb]
\centering
\includegraphics[width=\columnwidth]{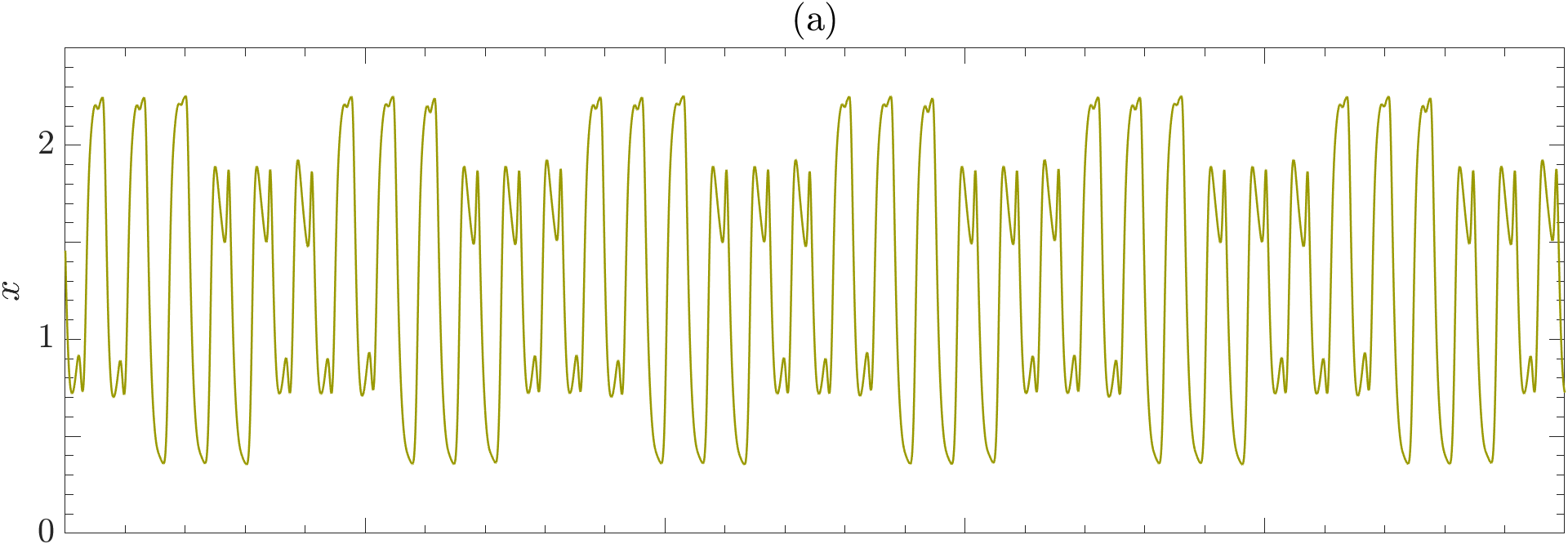}
\includegraphics[width=\columnwidth]{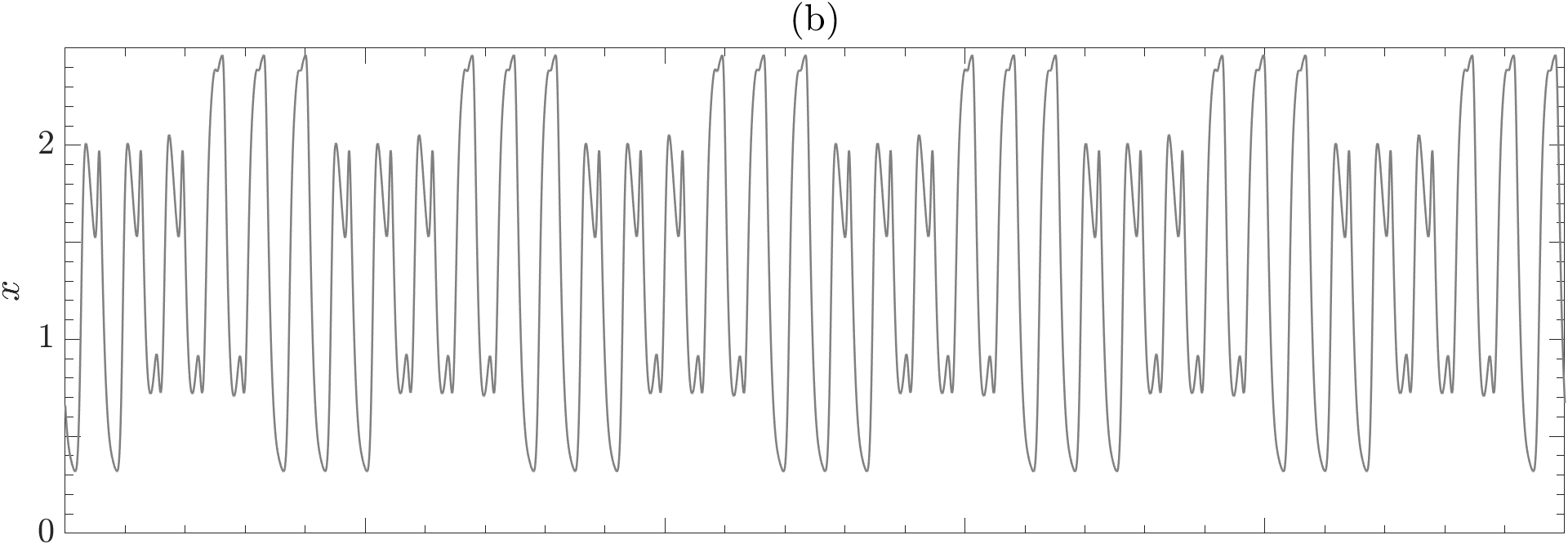}
\includegraphics[width=\columnwidth]{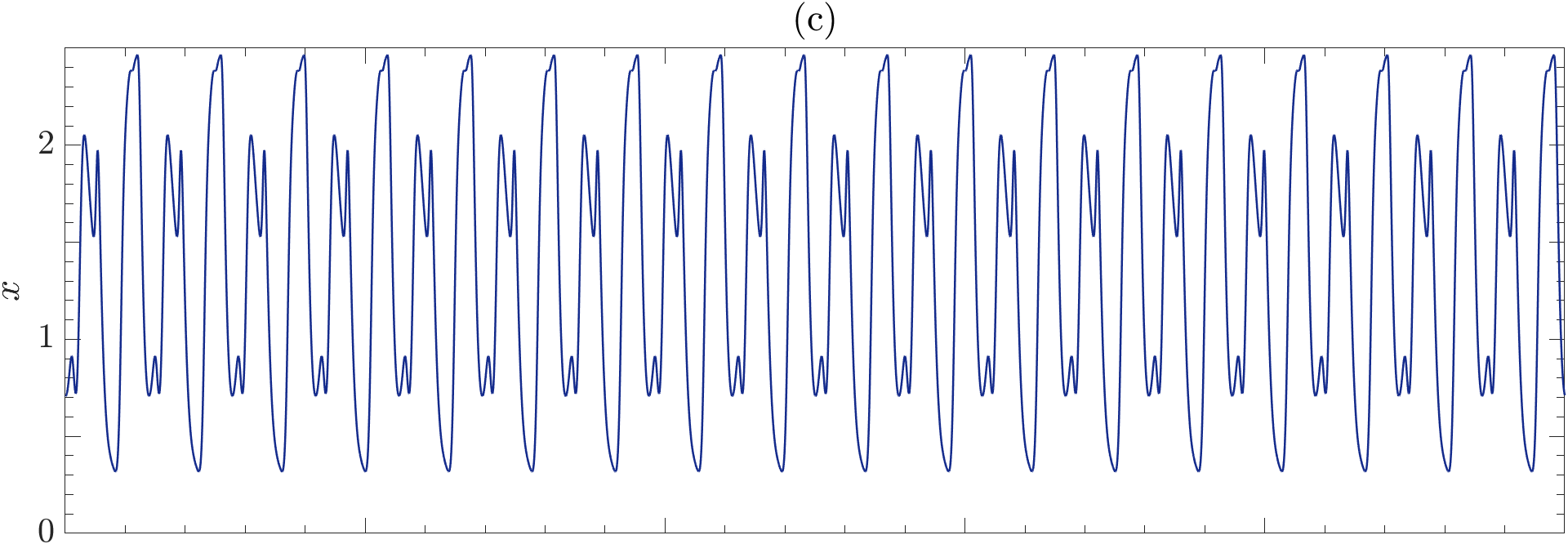}
\includegraphics[width=\columnwidth]{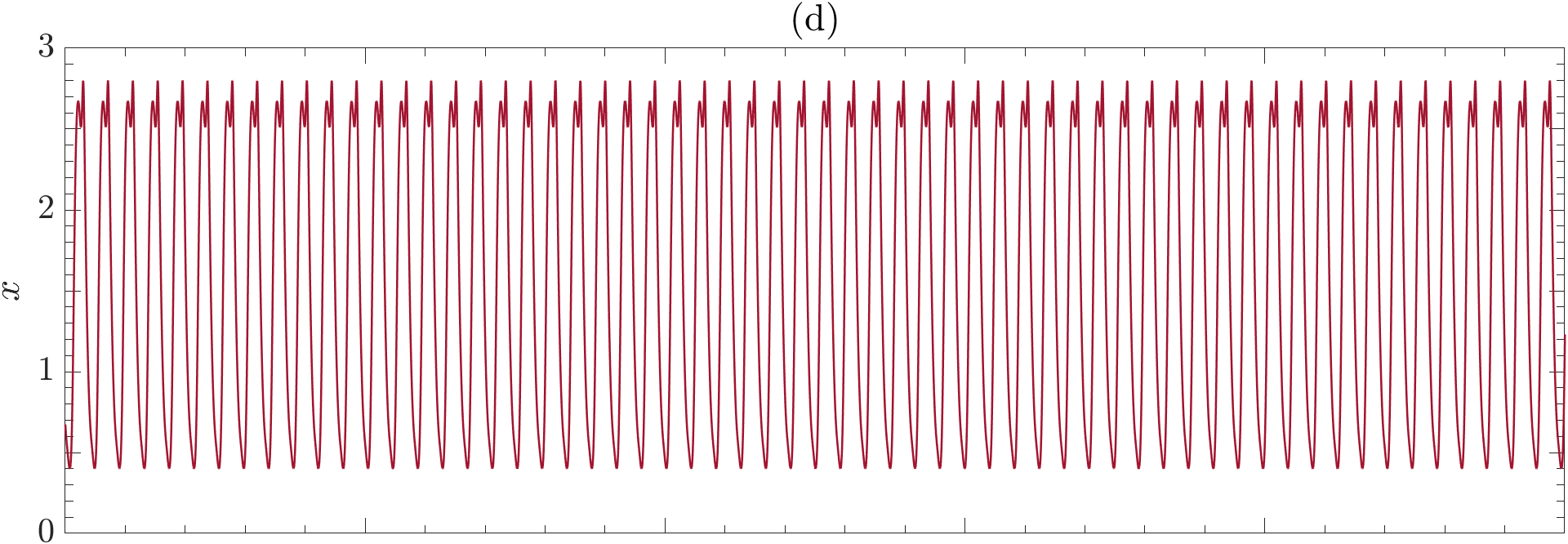}
\includegraphics[width=\columnwidth]{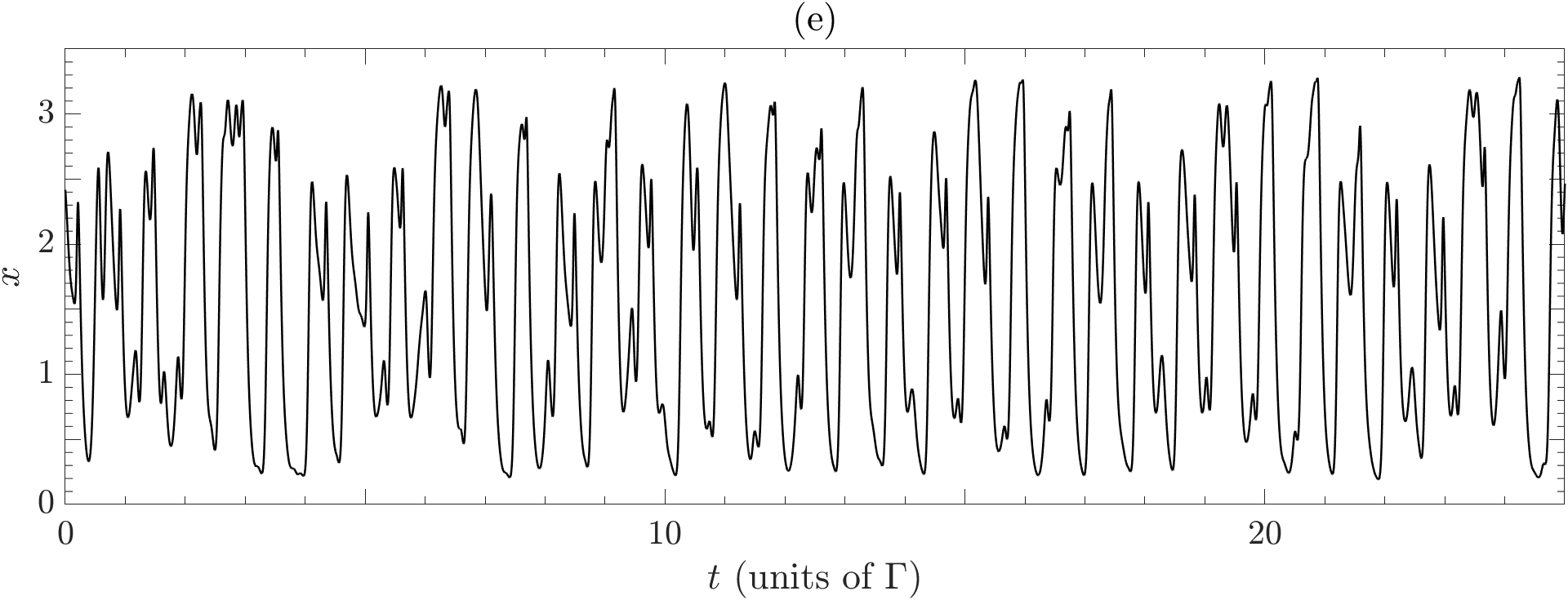}
\caption{Representative examples of time evolution of
  Eq. \eqref{eq::map}. The parameters correspond to $n = 4$ and
  $(\alpha, \Gamma) = $ $(4, 19.5)$ for (a), $(4.4, 17.2)$ for (b) and
  (c) and $(5.9, 19.7)$ for (d) and (e). Initial conditions correspond
  to Eq. \eqref{eq::init_cond} with parameters values $(A, x_{off}) =
  $ $(0.9, 1.01)$ for (a), $(0.2, 1.7)$ for (b), $(0.2, 1.69)$ for
  (c), $(0.9, 1.01)$ for (d) and $(0.5, 1.05)$ for (e). For all
  solutions a transient of $1000 \Gamma$ was discarded.}
\label{fig_timeSeries}
\end{figure}

To show the rich variety of behaviors found in the Mackey-Glass
system, in Fig. \ref{fig_timeSeries} we present examples of stationary
solutions of Eq. \eqref{eq::map} for different parameter values and
different initial conditions. Panels (a) through (d) show examples of
limit cycles while panel (e) shows an example of a chaotic
trajectory. We can see the multistability of the system as panels
(b)-(c) and (d)-(e) in Fig.~\ref{fig_timeSeries} correspond to the
same control parameter values but different initial conditions.

\begin{figure}[tb]
\centering
\includegraphics[width=0.49\columnwidth]{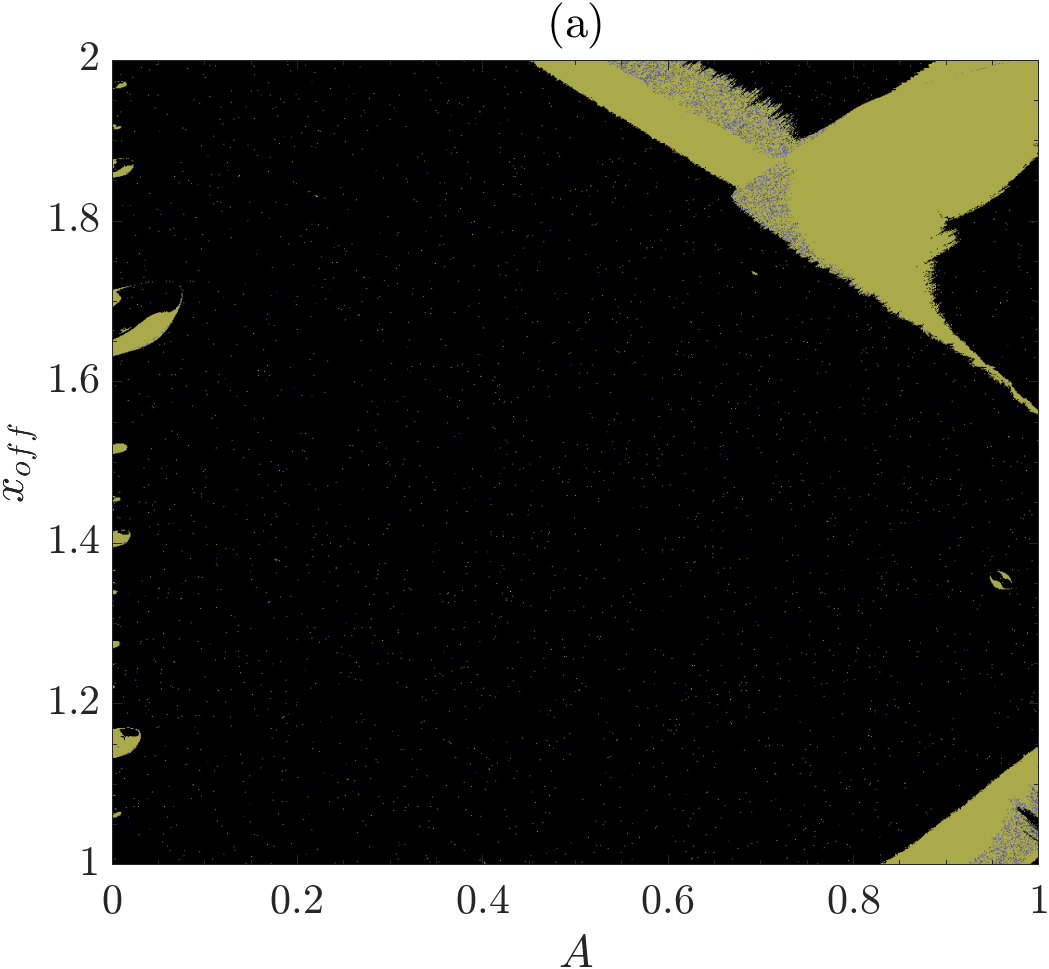}
\includegraphics[width=0.49\columnwidth]{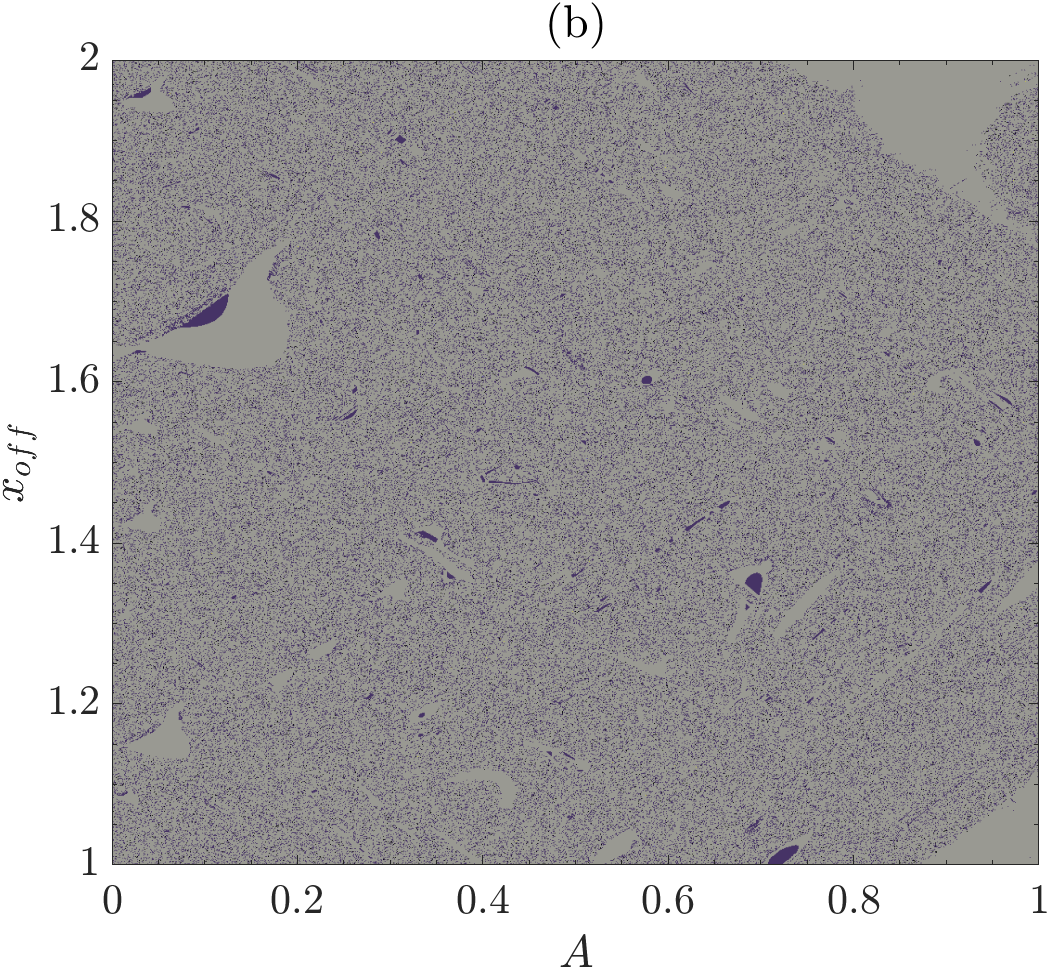}
\includegraphics[width=0.49\columnwidth]{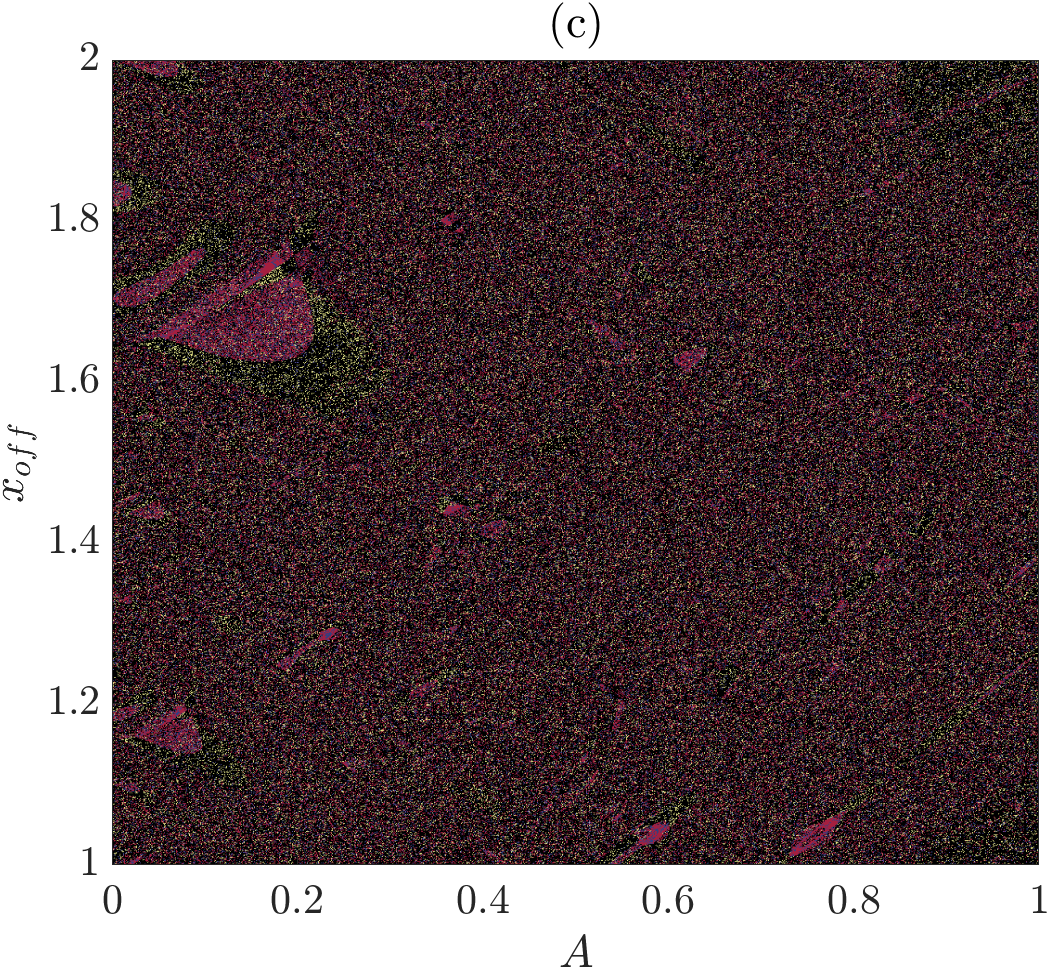}
\includegraphics[width=0.49\columnwidth]{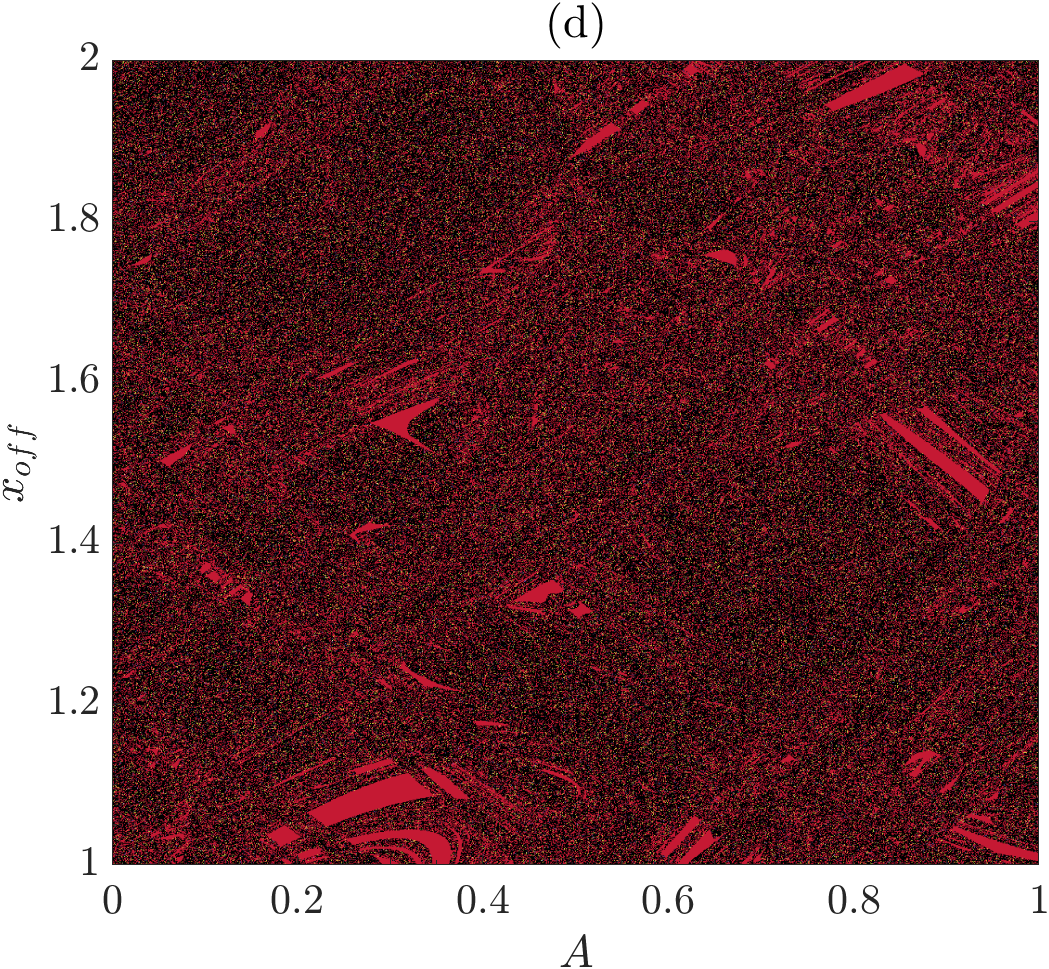}
\caption{Two-dimensional cross-sections of the state space of the
  system. Each color corresponds to a different final state of the
  system when using \eqref{eq::init_cond} as initial conditions. Black
  corresponds to chaotic behavior while color indicates a limit cycle
  attractor (using the same code as in
  Fig.~\ref{fig_timeSeries}). Parameters are $(\alpha, \Gamma) = $
  $(4, 19.5)$ for panel (a), $(4.4, 17.2)$ for panel (b), $(4.6,
  18.2)$ for panel (c) and $(5.9, 19.7)$ for panel (d).}
\label{fig_phaseSpace}
\end{figure}

To further study the multistability of the system we plotted relevant
cross-sections of the systems state space. Figure \ref{fig_phaseSpace}
illustrates the attractors reached for points of two-dimensional
cross-sections of the state space of the system and four different
pairs of parameters $\Gamma$ and $\alpha$.  The color code used in
Fig.~\ref{fig_phaseSpace} was chosen with the same criteria as for
Fig.~\ref{fig_timeSeries}. The cross-sections correspond to the
following parametric function of initial conditions
\begin{equation}
    x_i^{in} = A\sin(2\pi t_i) + x_{off}, \quad t_i = -\frac{N-i+1}{N}\Gamma
    \label{eq::init_cond}
\end{equation}
where $A$ and $x_{off}$ are the parameters of this family of
functions.  In the diagrams of Fig.~\ref{fig_phaseSpace}, we can
appreciate the structural richness exhibited by the space of initial
conditions. As the system's parameters change, both the number of
solutions and the nature of the boundaries separating the basins of
attraction vary. In some regions, the boundaries are relatively
smooth, while in others, they become increasingly intermixed as the
parameters change. It is also evident that, for certain parameter
values, different periodic solutions coexist, and in others, chaotic
trajectories emerge.

This approach to studying multistability in delayed systems, as
explored in Refs.  \onlinecite{amil2015organization,
  tarigo2022characterizing, tarigo2024basin}, offers the advantage of
visualizing the basins of attraction of a high-dimensional system
within a two-dimensional diagram. However, it has the drawback of
being highly dependent on the specific cross-sections chosen or the
family of functions used. To address this limitation, in the next
section, we randomly sample the state space to count the different
solutions of the system and quantify their relative importance.

\section{Characterizing multistability: Counting solutions and basin entropy} \label{sec:results:qm}

\begin{figure}[tb]
\centering
\includegraphics[width=0.495\columnwidth]{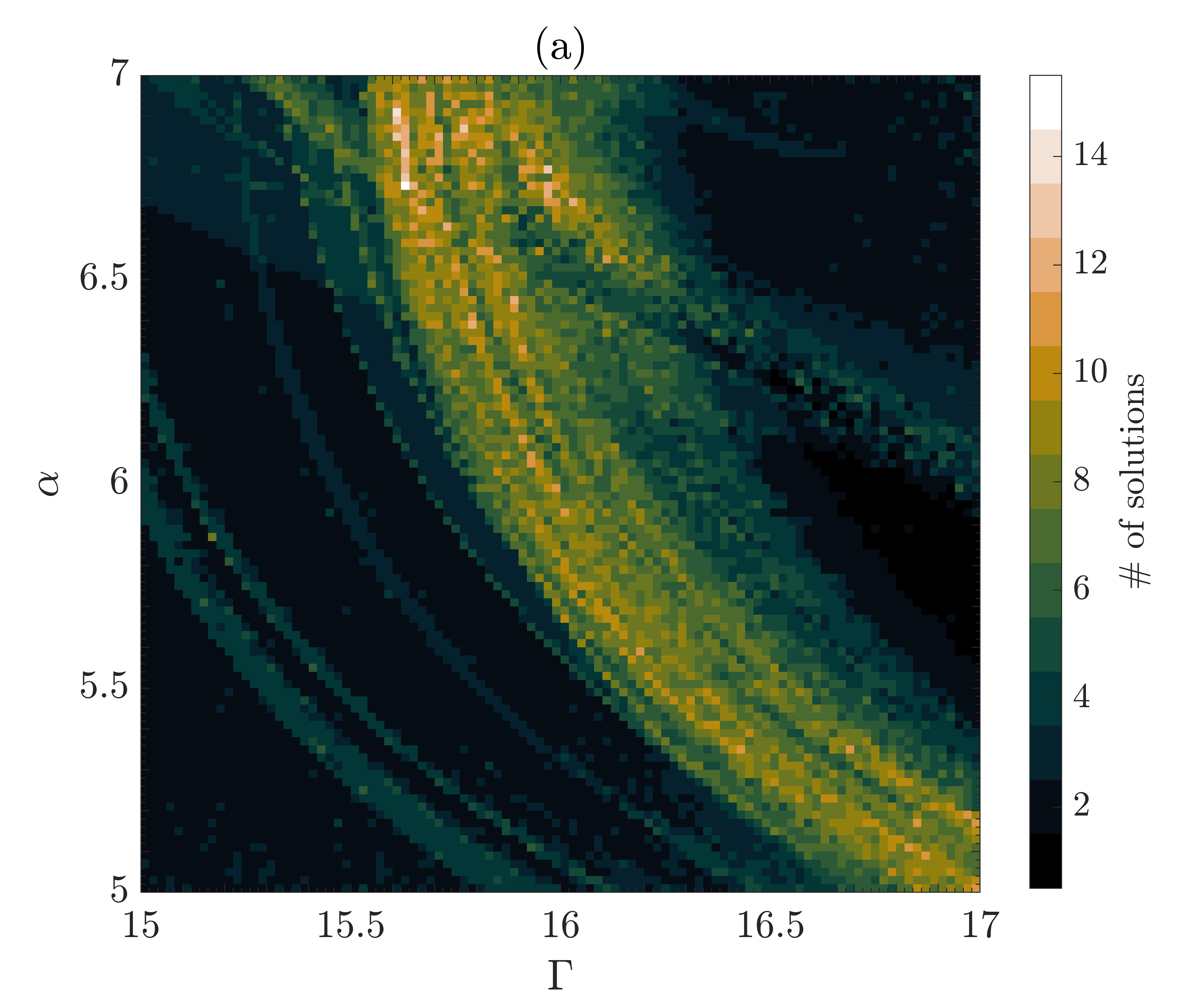}
\includegraphics[width=0.495\columnwidth]{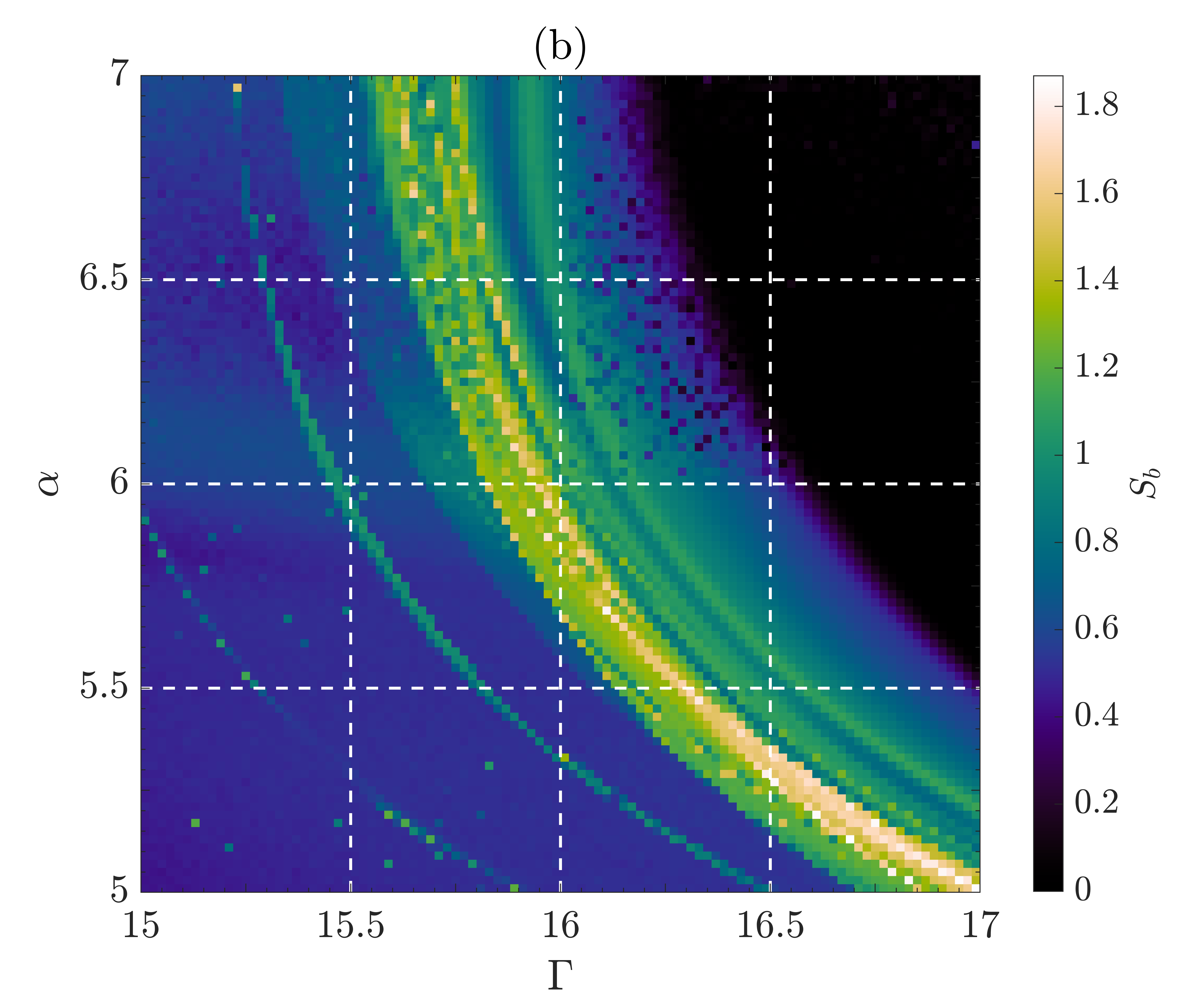}
\includegraphics[width=0.495\columnwidth]{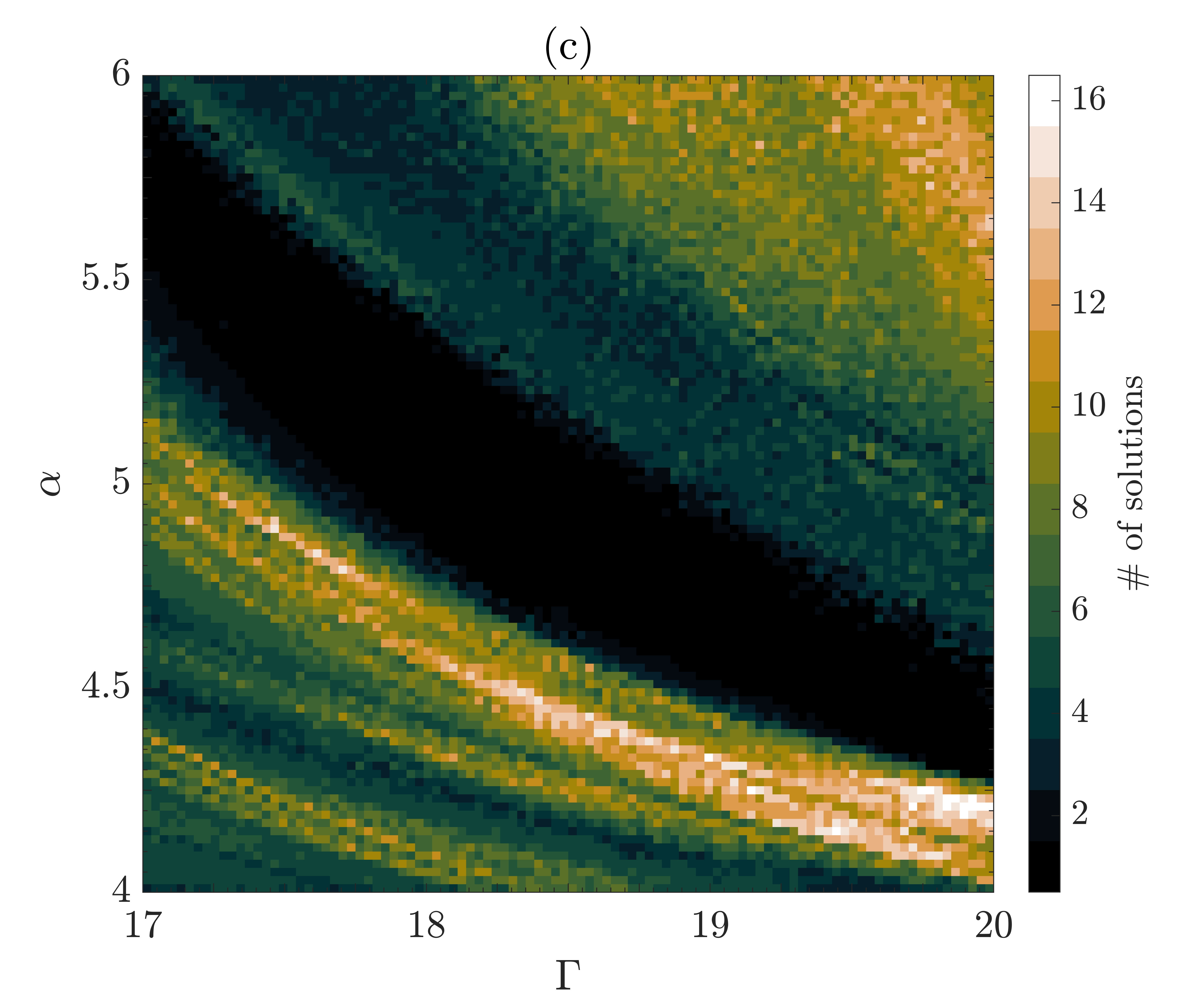}
\includegraphics[width=0.495\columnwidth]{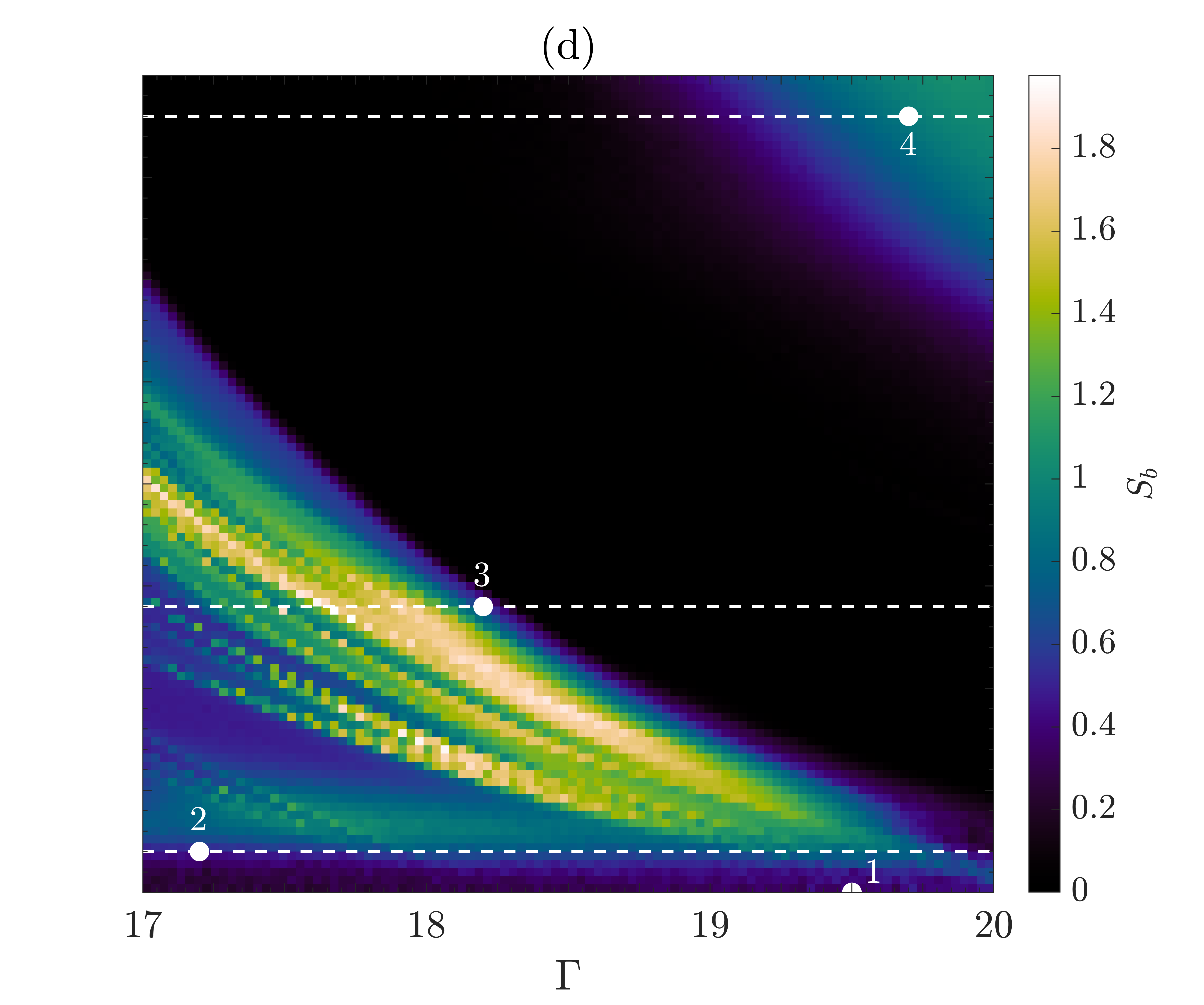}
\caption{Multistability in the parameter space of the system. Panels
  (a) and (c) show the number of distinct attractors for each point in
  the parameter space while panels (b) and (d) show its corresponding
  basin entropy. For each point in the parameter space $10^5$ initial
  conditions were sampled randomly from a uniform distribution between
  $0$ and $2$ evenly distributed between $100$ boxes. Then, solutions
  were matched through its period, its number of maxima per period and
  the ordering of those maxima in order to obtain the number of
  different attractors.  Points labelled $1,2,3$ and $4$ correspond to
  the parameter values chosen in panels a,b,c and d (respectively) of
  Fig.~\ref{fig_phaseSpace} while dashed lines correspond to the
  diagrams of Fig.~\ref{fig_fraction}.}
\label{fig_be}
\end{figure}

In order to have a first approach to the way in which the basins of
attraction are organized we counted the different coexisting solutions
sampling the state space randomly and compared with the calculated the
basin entropy using the stochastic method described in
Sec.~\ref{sec:be}.  Figure \ref{fig_be} display a general overview of
the number of attractors (left panels) and the basin entropy (right
panels) as a function of the control parameters, $\Gamma$ and
$\alpha$. Panels (a) and (c) show the quantity of distinct solutions
in the parameter space, sampling initial conditions randomly. To
achieve this, we sampled $10^5$ initial conditions from a uniform
distribution between $0$ and $2$ in contrast to the cross-sectional
method explained earlier. For the evolution, we discarded a transient
of $1000 \Gamma$ and used a $200 \Gamma$ long time series to classify
the different attractors. We calculated the period, the number of
maxima per period and the ordering of those maxima for each time
series to characterize and distinguish each attractor.

As we can see in Fig.~\ref{fig_be}(a) and (c) the system is highly
multistable, exhibiting from only one up to 16 coexisting solutions in
some of the regions explored. This fact raises the question about the
nature of the corresponding basins of attraction for each solution and
the characteristics of the interface that separates them.  To answer
these questions Fig.~\ref{fig_be}(b) and (d) show the corresponding
basin entropy of each region calculated using the method described in
section \ref{sec:be}. The basin entropy also reveals interesting
dynamics in Fig~\ref{fig_be}(b) and (d), in both panels there appears
to be a curve where the basin entropy exhibits a local maxima.  In the
next Section we will show that this curve is near the extinction of
the limit-cycle solutions or at least the dominance of the chaotic
attractor (see Fig.~\ref{fig_fraction}).  This behavior of the basin
entropy indicates that the structure of the basins of attraction
becomes very intricate close to this curve and could the indicator of
a bifurcation \cite{daza2023unpredictability, tarigo2024basin}.

Figure \ref{fig_be} also demonstrates that basin entropy does not
necessarily correlate with a large number of coexisting
attractors. For instance, the bottom right corner of
Fig.~\ref{fig_be}(c) exhibits the highest number of coexisting
attractors, yet the entropy in that region is nearly zero. The points
in Fig.~\ref{fig_be}(d) correspond to the cross sections in
Fig.~\ref{fig_phaseSpace}. To explore this further, in the next
Section we measure the evolution of the basin fraction for each
attractor as the parameters were varied along the dashed lines shown
in Fig.~\ref{fig_be}(b) and (d).

\section{Relationship between basin entropy and basin fraction} \label{sec:results:bf}

\begin{figure*}[tb]
\centering

\includegraphics[width=0.33\textwidth]{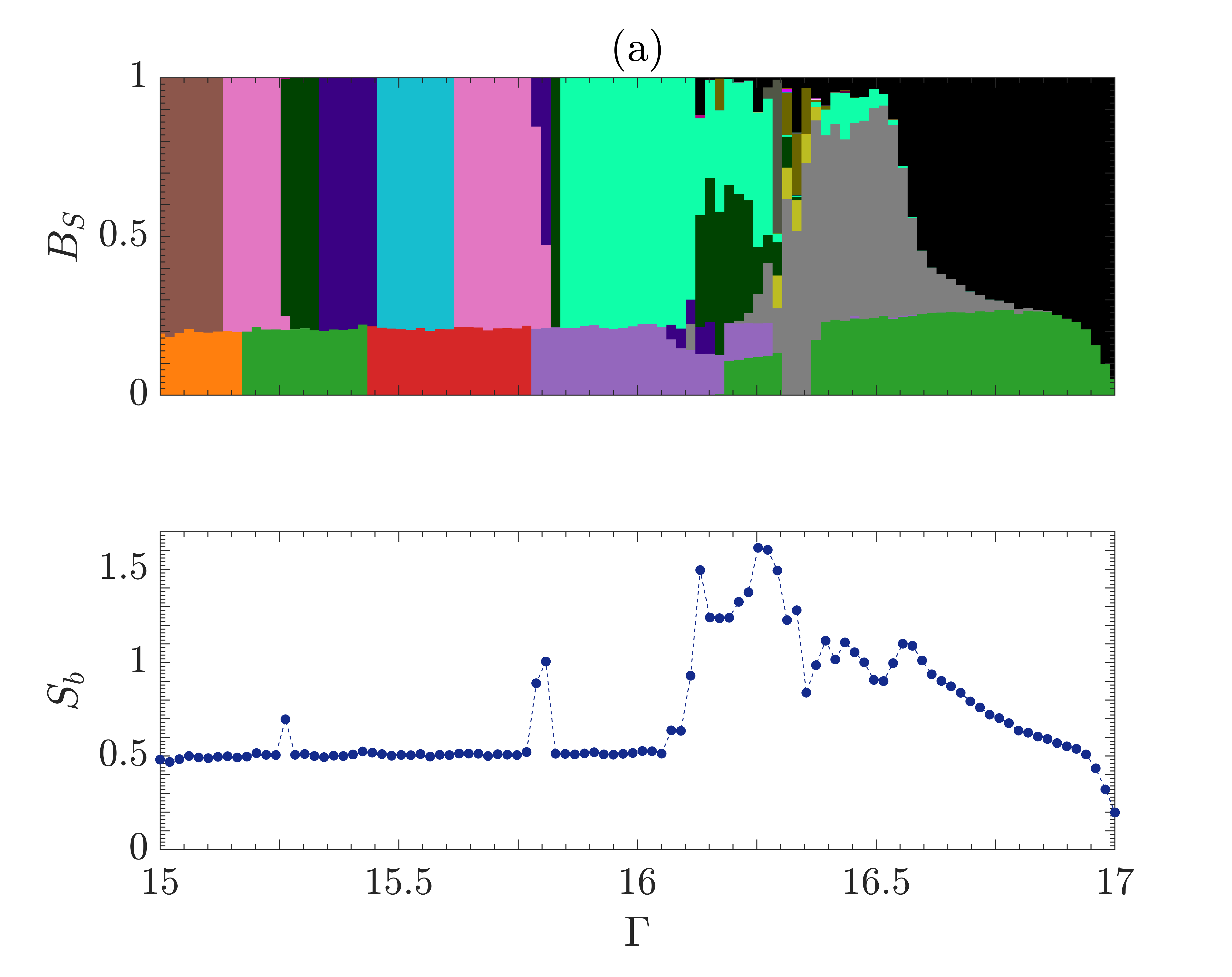}
\includegraphics[width=0.33\textwidth]{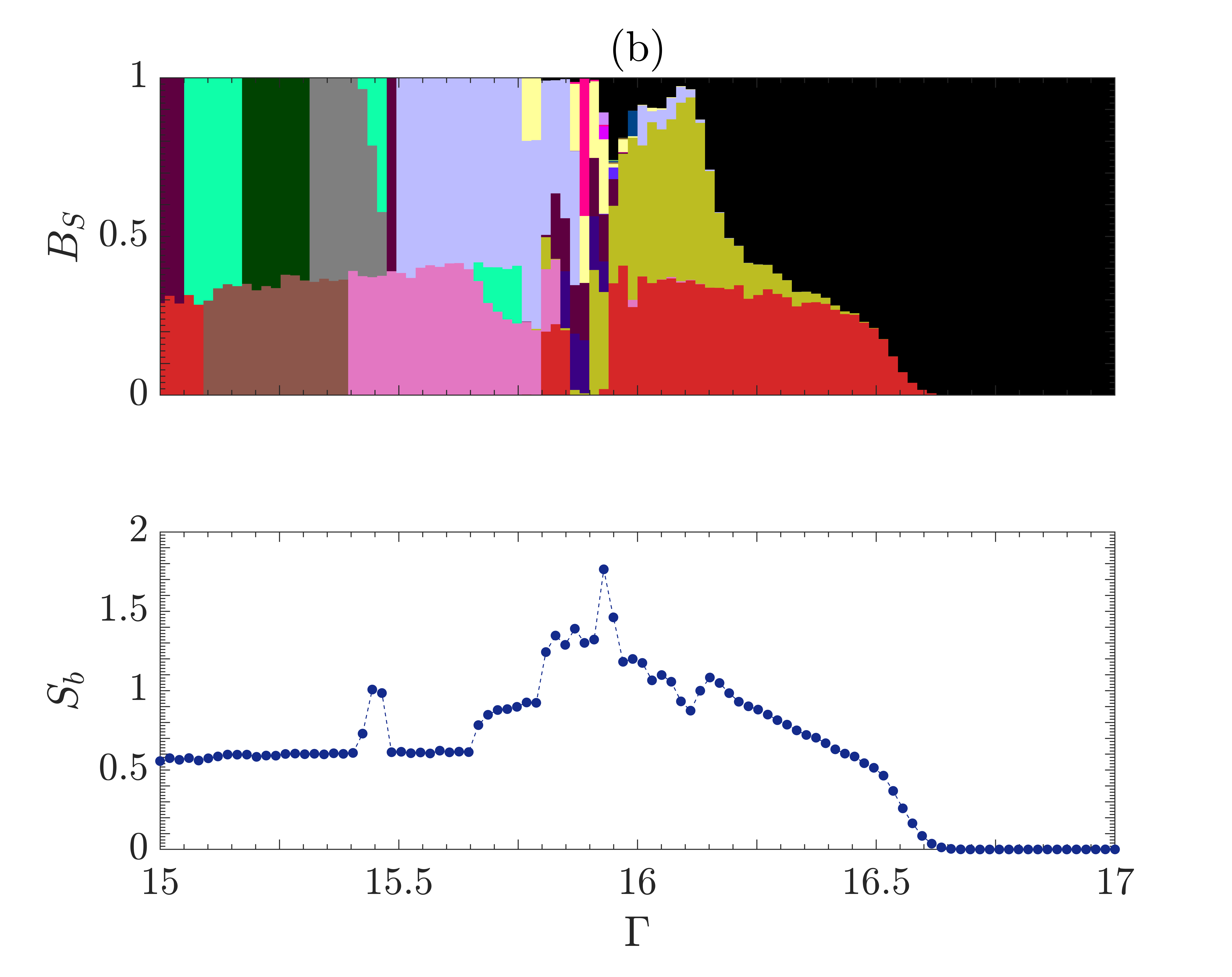}
\includegraphics[width=0.33\textwidth]{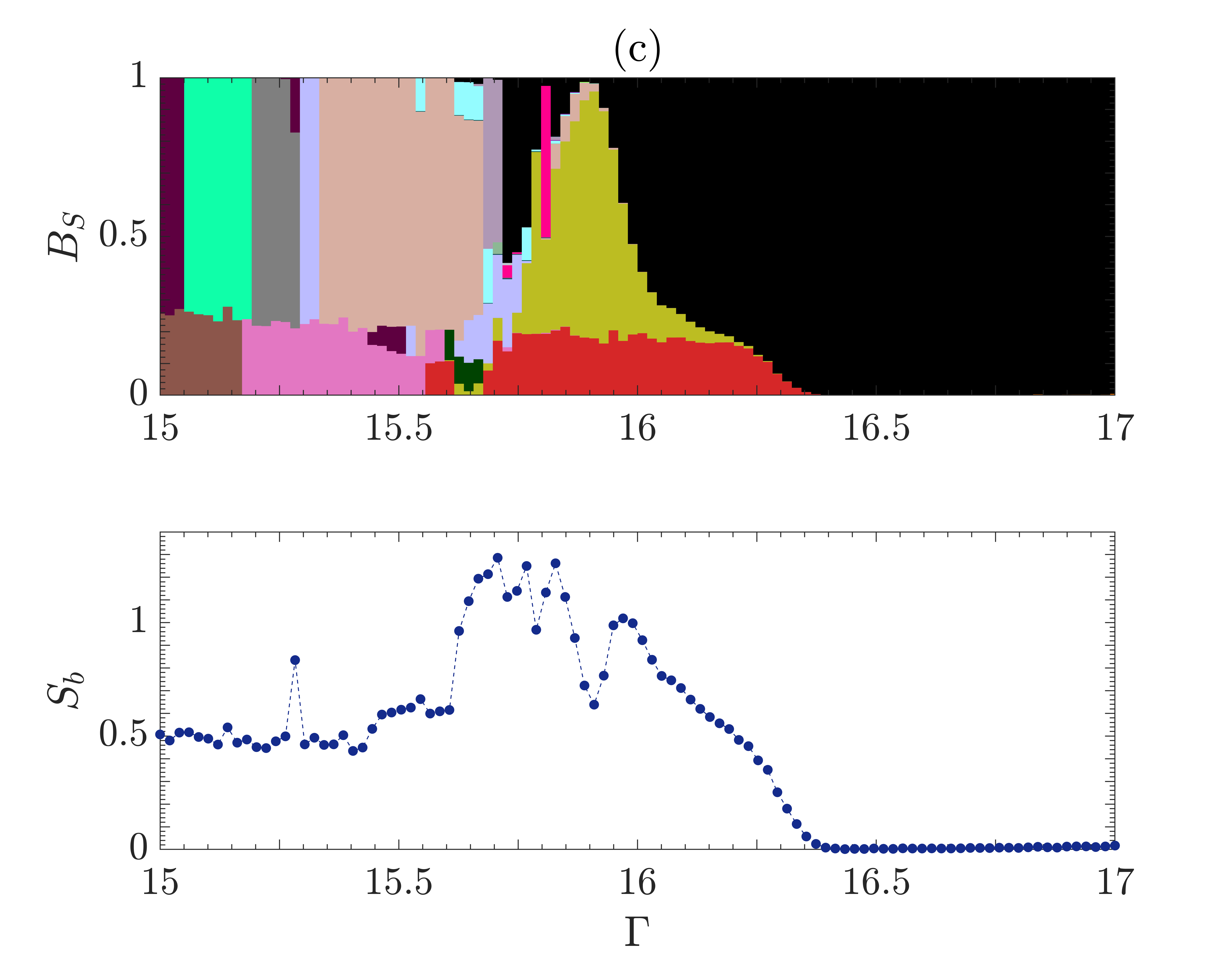}

\includegraphics[width=0.33\textwidth]{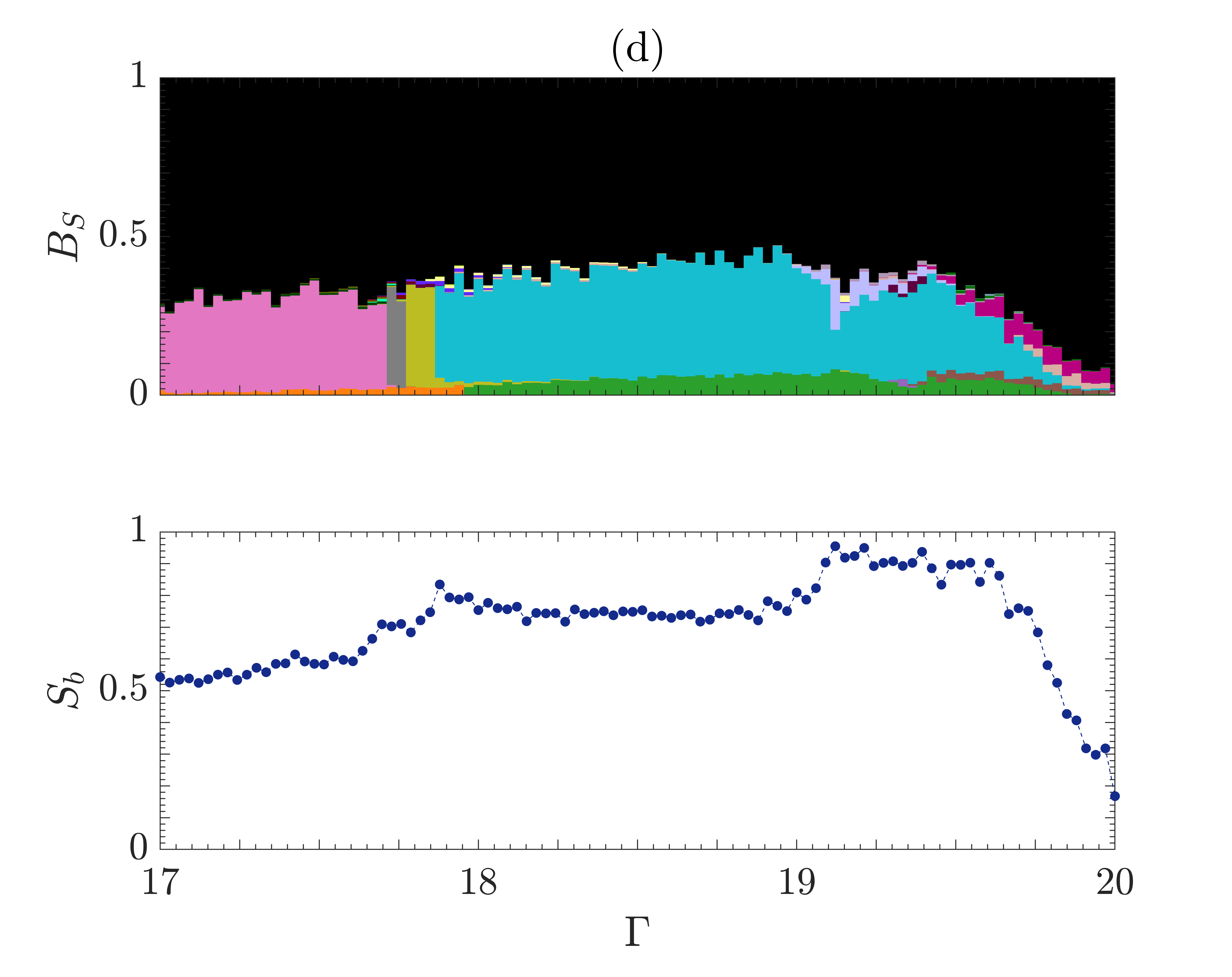}
\includegraphics[width=0.33\textwidth]{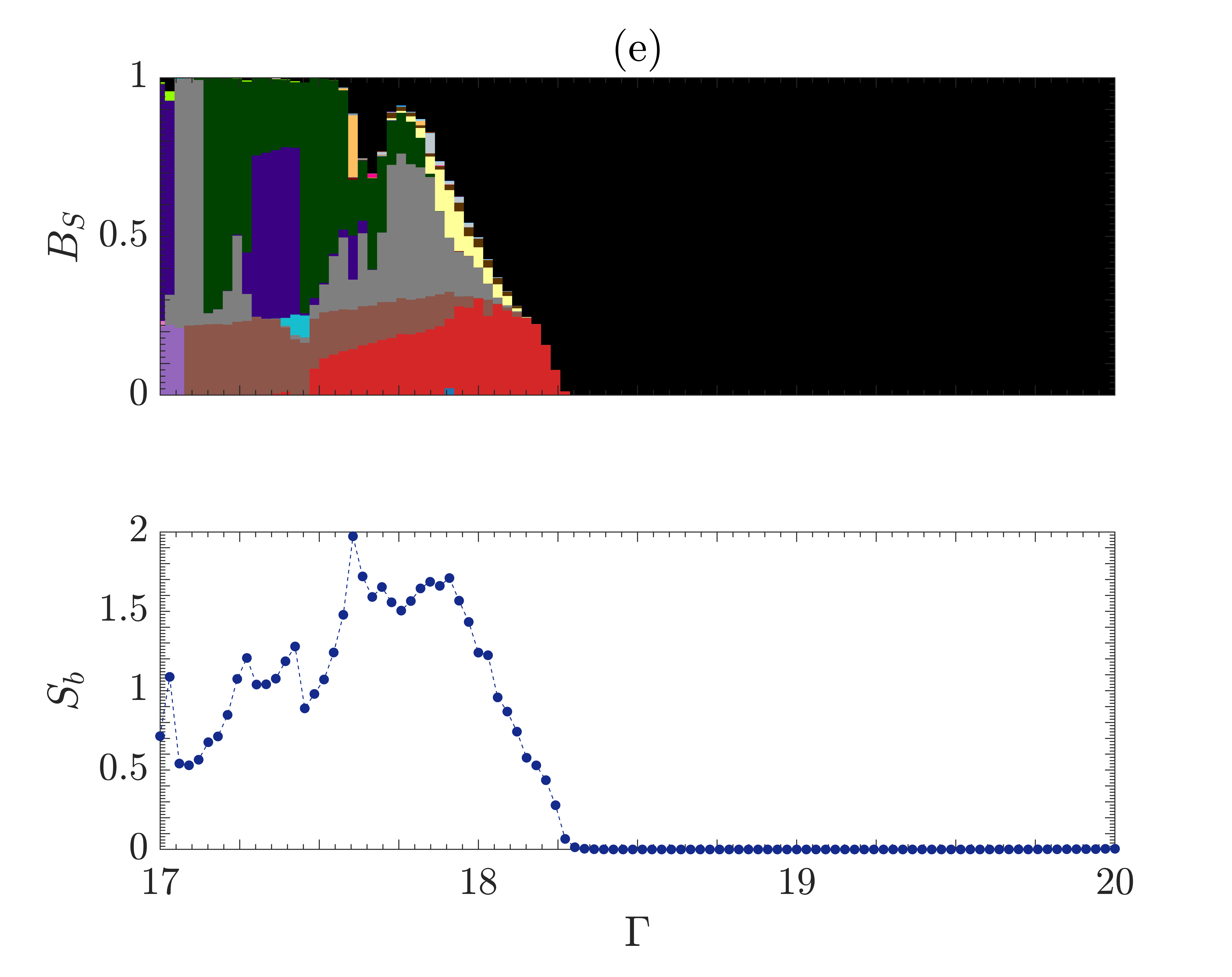}
\includegraphics[width=0.33\textwidth]{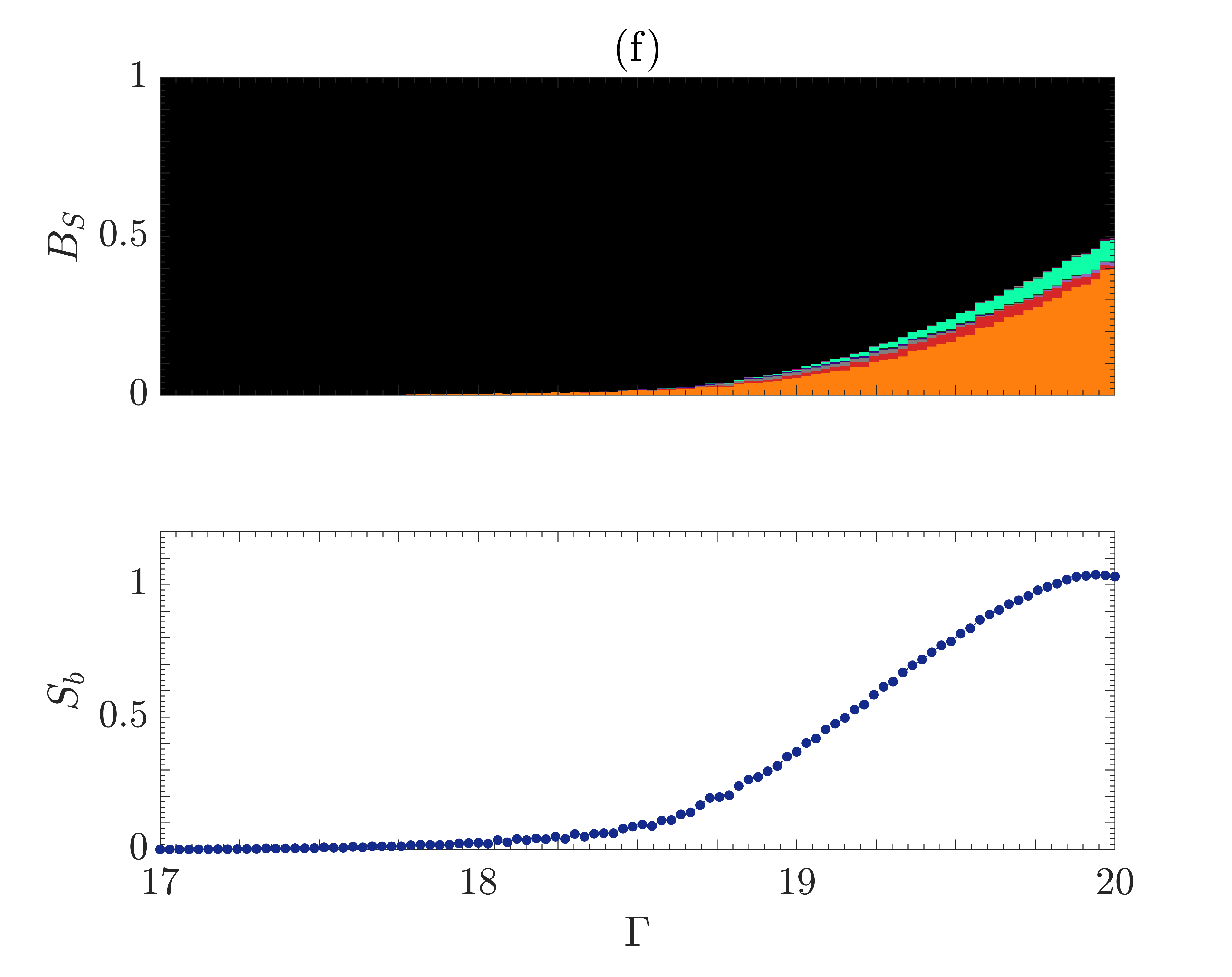}

\includegraphics[width=0.33\textwidth]{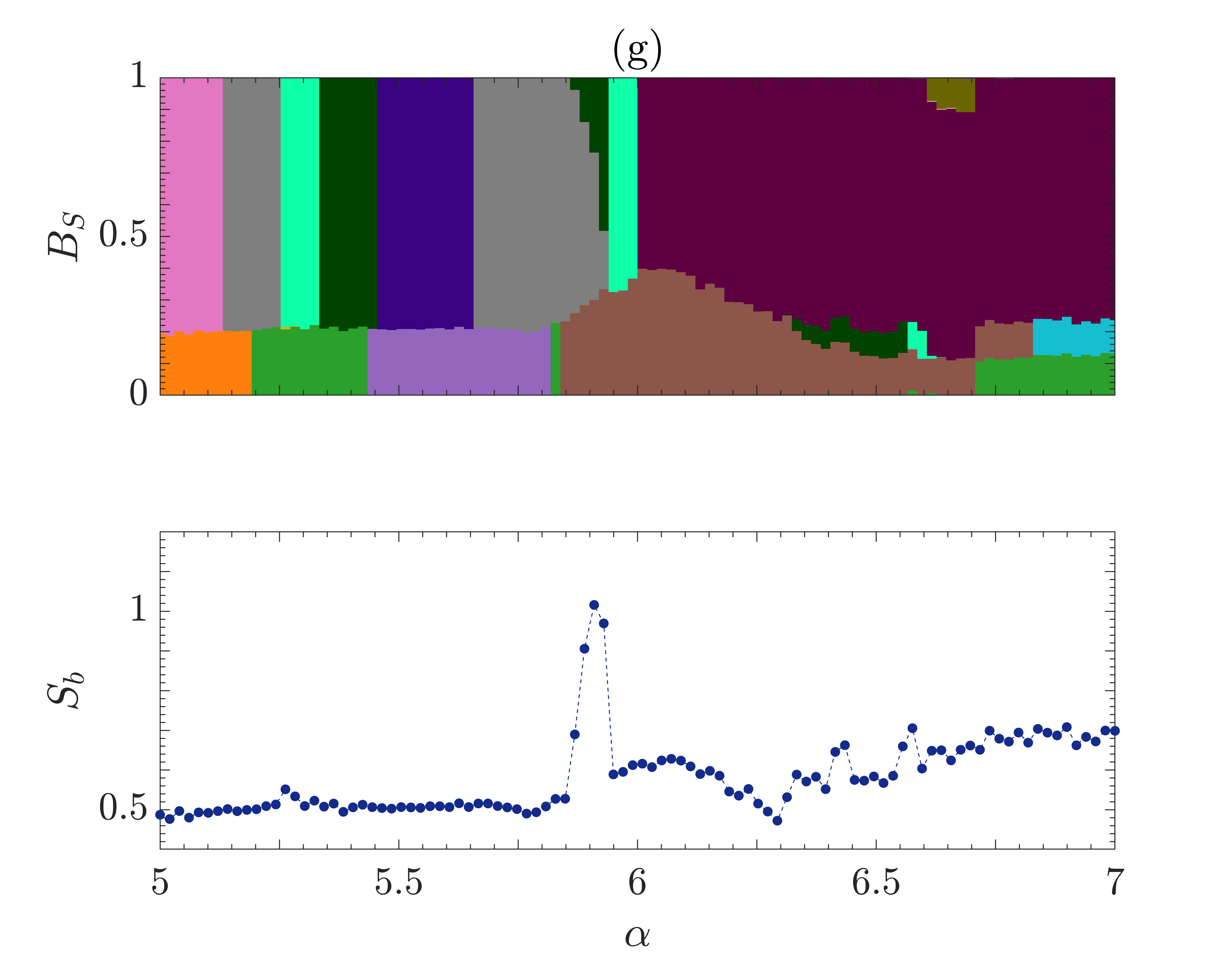}
\includegraphics[width=0.33\textwidth]{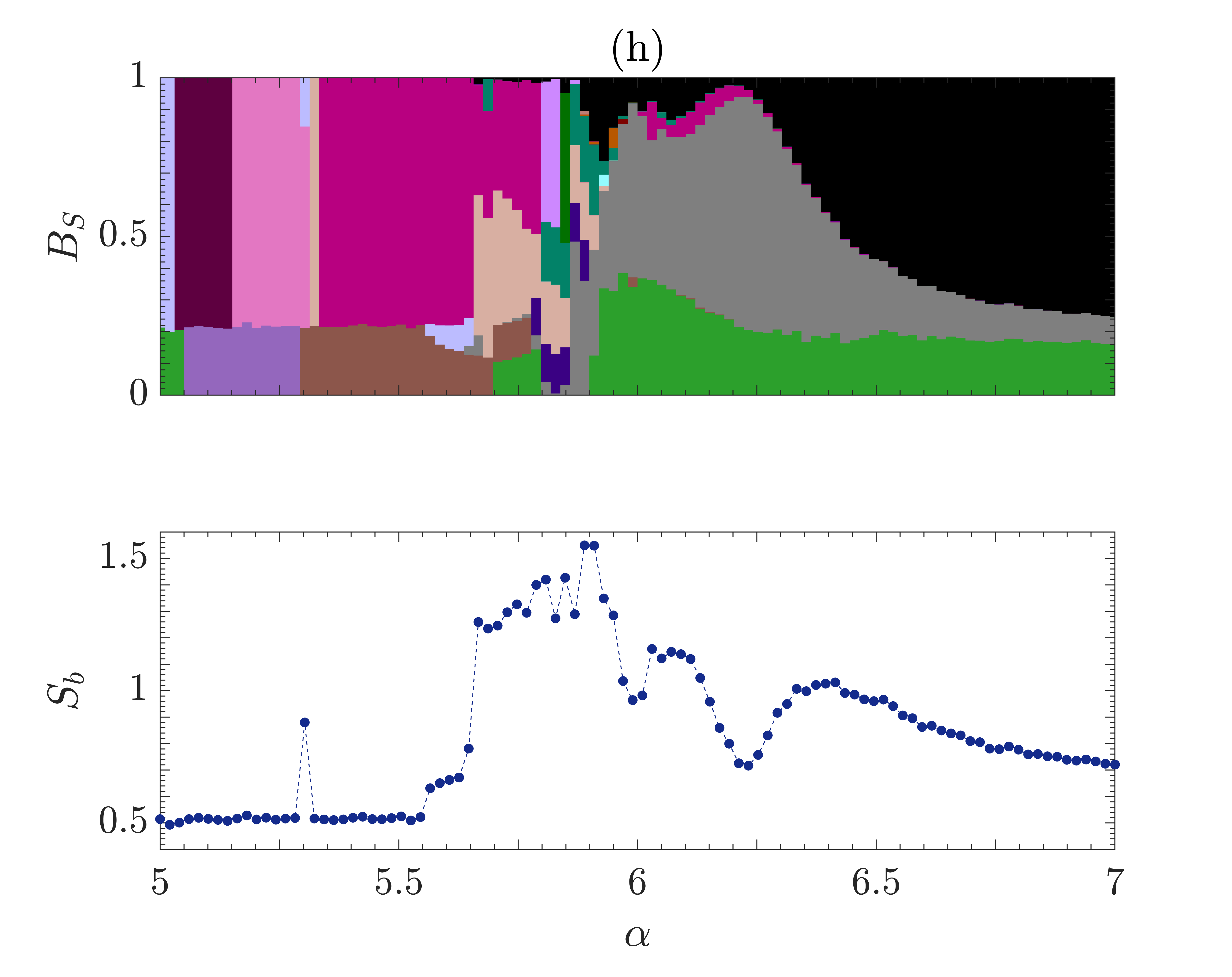}
\includegraphics[width=0.33\textwidth]{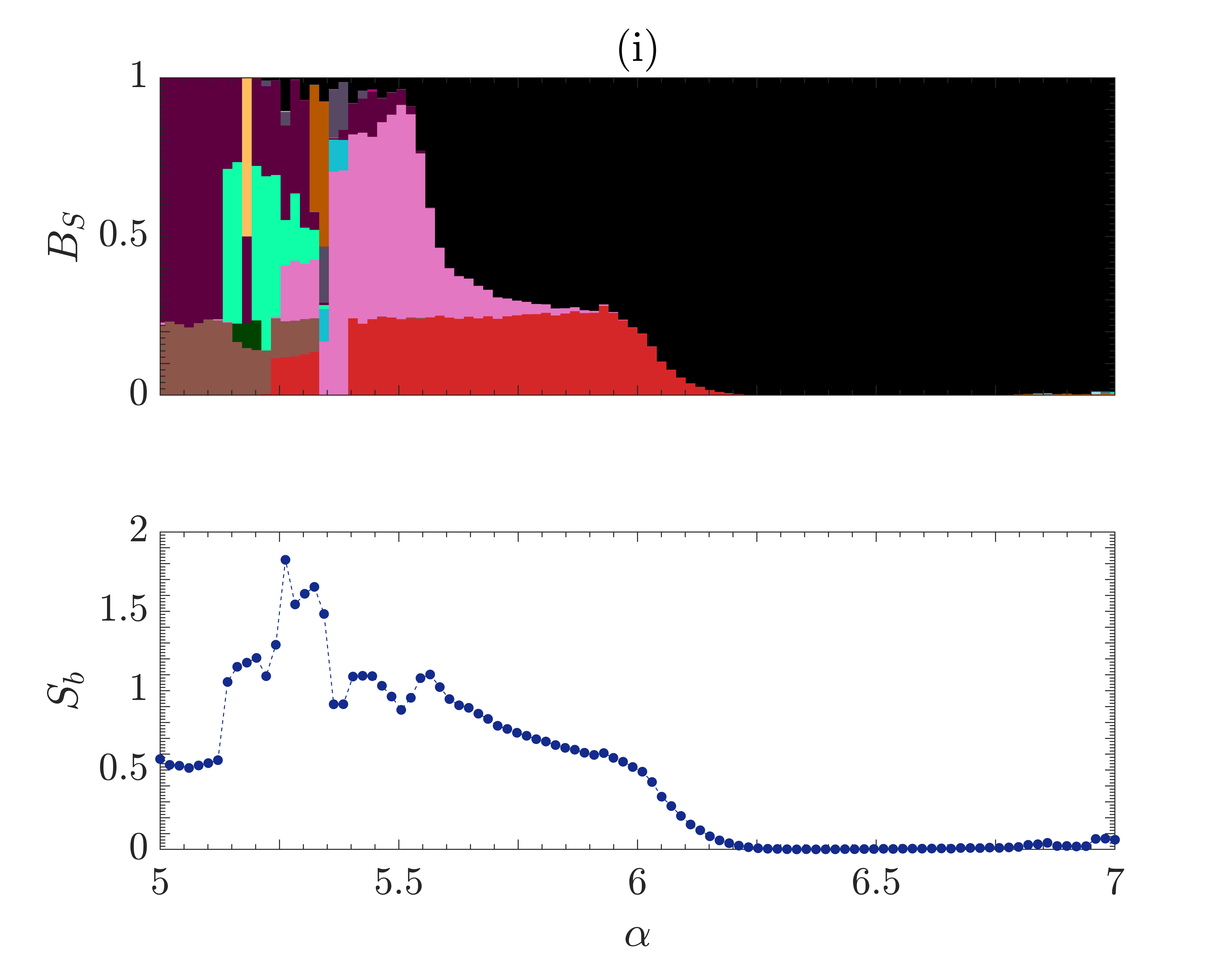}

\caption{Basin fraction (top) and basin entropy (bottom) for 9 lines
  in the parameter space. Black corresponds to chaotic behavior while
  color indicates a limit cycle attractor. (a), (b) and (c) correspond
  to $\Gamma \in [15, 17]$ and $\alpha = 5.5$, $\alpha = 6$ and
  $\alpha = 6.5$ respectively, (d), (e) and (f) correspond to $\Gamma
  \in [17, 20]$ and $\alpha = 4.1$, $\alpha = 4.7$ and $\alpha = 5.9$
  respectively and (g), (h) and (i) correspond to $\alpha \in [5, 7]$
  and $\Gamma = 15.5$, $\Gamma = 16$ and $\Gamma = 16.5$
  respectively.}
\label{fig_fraction}
\end{figure*}

To better understand the structure of basins of attraction in
Fig.~\ref{fig_fraction}, we jointly show the fraction of basins and
the entropy of basins along sections of parameter space of constant
$\Gamma$ or $\alpha$. We can observe that as the parameters change,
different attractors emerge and others disappear and their basin
fraction and basin entropy also varies. We can also notice that
changes in the basin entropy do not necessarily correlate with changes
in the basin fraction. The reason for that is that even though the
basins of attraction may occupy a larger or smaller fraction their
basin boundaries may or may not become more or less riddled.

Figure \ref{fig_fraction}(d) reveals why the basin entropy is low in
the bottom right corner of Fig~\ref{fig_be}(d) even though there is
very high number of coexisting attractors in the bottom right corner
of Fig.~\ref{fig_be}(c). Most attractors in this region of the
parameters space have very low basin fraction compared to the strange
attractor meaning they occupy a small region of the state space and
thus their contribution to the basin entropy is negligible. So even
though there is a very high attractor count in this region the strange
attractor dominates the state space rendering basin entropy almost
zero.

As we can see in some regions of Fig. \ref{fig_fraction} the basin
entropy remains constant but the attractors measured change (different
colors in the basin fraction diagrams) even though there is no change
in the basin fraction. In other instances there is a spike in basin
entropy where the attractor becomes a new one. The fact that basin
entropy and basin fraction remain constant even when changing
attractor is an indication of an immutability of the basins of
attraction in those regions even though the attractors themselves are
experiencing some mutations. Moreover, the spikes in the basin entropy
indicate that for those parameters not only the attractors mutate but
their basins of attraction get more intermixed. This results could be
hinting at bifurcations in the system and could be further studied
using attractor continuation as in
Ref. \onlinecite{datseris2023framework} to assess if the difference
when the basin entropy remains constant are just continuations of the
old ones or if they are actually new attractors.

\section{Conclusions} \label{sec:conc}
In this work, we investigated the structure of the
infinite-dimensional initial condition space of the Mackey-Glass
system using complementary techniques.  We generalized the concept of
basin entropy by randomly sampling the points where we centered the
boxes defining set of initial conditions and showed that this method
converges rapidly to the basin entropy calculated taking regular
lattices. This generalization allowed us to obtain representative
samples of the infinite-dimensional space and gain a global
understanding of how the basins of attraction are distributed. This
method was complemented by counting the number of distinct basins and
determining the fraction of volume occupied by each one.

We applied this new method to a discretization of the Mackey-Glass
delayed model, which has been previously validated through
experimental and numerical approaches. Our results show that, when
combined with the basin fraction, this method can enhance our
understanding of the composition of basins of attraction in
high-dimensional systems. We found that the system's unpredictability,
arising from its multistability, is influenced not only by the number
of distinct coexisting attractors but also by the relative volume of
their corresponding basins and the morphology of their boundaries.
The application of these techniques to other high-dimensional
dynamical systems could offer deeper insights into their
multistability and predictability, potentially uncovering new
relationships between basin structures, system dynamics, and
underlying bifurcations. Given that basin entropy is still a
relatively new tool, its potential applications are actively
expanding, as demonstrated by recent research. We believe that our
proposed modification—stochastically sampling the space of initial
conditions—extends its utility to a broader range of high-dimensional
systems of interest across various disciplines, particularly those
with feedback or delay dynamics.

\begin{acknowledgments}We thank Cristina Masoller (UPC, Spain) for the stimulating discussions.
We acknowledge support from the \textit{Programa de desarrollo de las
  Ciencias Básicas} (PEDECIBA), MEC-Udelar, (Uruguay). The numerical
experiments were performed at the ClusterUY (site: https://cluster.uy)
\end{acknowledgments}

%
\end{document}